\documentclass[aps,twocolumn,epsfig,rotate,superscriptaddress,showpacs]{revtex4}
\usepackage[dvips]{graphics}
\usepackage{graphicx}
\usepackage{amsfonts}
\usepackage{amssymb}
\usepackage{amsmath}
\usepackage{subfigure}
\usepackage{prettyref}
\usepackage{float}
\usepackage{amsmath}
\usepackage{amssymb}
\usepackage{esint}
\begin{document}

%\singlespacing

\title{Universal Dynamical Decoupling: Two-Qubit States and Beyond}
\author{Musawwadah Mukhtar}
\affiliation{Department of Physics, National University of
Singapore, 117542, Republic of Singapore}
\author{Thuan Beng Saw}
\affiliation{Department of Physics, National University of Singapore, 117542, Republic of Singapore}
\author{Wee Tee Soh}
\affiliation{Department of Physics, National University of Singapore, 117542, Republic of Singapore}
\author{Jiangbin Gong}
\email[]{phygj@nus.edu.sg}
\affiliation{Department of Physics, National University of Singapore, 117542, Republic of Singapore}
\affiliation{Centre for Computational Science and Engineering,
\\
National University of Singapore, 117542, Republic of Singapore}
\affiliation{NUS Graduate School for Integrative Sciences and Engineering, Singapore
117597, Republic of Singapore}
\date{\today}

\begin{abstract}
Uhrig's dynamical decoupling pulse sequence has emerged as one
universal and highly promising approach to decoherence suppression.
So far both the theoretical and experimental studies have examined
single-qubit decoherence only. This work extends Uhrig's universal
dynamical decoupling from one-qubit to two-qubit systems and even to
general multi-level quantum systems. In particular, we show that by
designing appropriate control Hamiltonians for a two-qubit or a
multi-level system, Uhrig's pulse sequence can also preserve a
generalized quantum coherence measure to the order of
$1+O(T^{N+1})$, with only $N$ pulses.  Our results lead to a very
useful scheme for efficiently locking two-qubit entangled states.
Future important applications of Uhrig's pulse sequence in
preserving the quantum coherence of multi-level quantum systems can
also be anticipated.
\end{abstract}

\pacs{03.67.Pp, 03.65.Yz, 07.05.Dz, 33.25.+k}
\maketitle

\section{Introduction}

Decoherence, i.e., the loss of quantum coherence due to system-environment coupling,
is a major obstacle for a variety of fascinating quantum information tasks. Even with the assistance of error corrections,
decoherence must be suppressed below an acceptable level to realize a useful quantum operation.
Analogous to refocusing techniques in nuclear magnetic resonance (NMR) studies, the dynamical decoupling (DD) approach to decoherence suppression
has attracted tremendous interest. The central idea of DD is to use a control pulse sequence to effectively decouple a quantum system from its environment.
%MM

During the past years several DD pulse sequences have been proposed.
The so-called ``bang-bang" control has proved to be very useful
\cite{BangBang1,BangBang2,exp2} with a variety of extensions.  However, it is not optimized
for a given period $T$ of coherence preservation. The
Carr-Purcell-Meiboom-Gill (CPMG) sequence from the NMR context can
suppress decoherence up to $O(T^{3})$ \cite{UDDexact}. In an approach called ``concatenated
dynamical decoupling" \cite{CDD1,CDD2}, the decoherence can be
suppressed to the order of $O(T^{N+1})$ with $2^{N}$ pulses.
Remarkably, in considering a single qubit subject to decoherence
without population relaxation, Uhrig's (optimal) dynamical
decoupling (UDD) pulse sequence proposed in 2007 can suppress
decoherence up to $O(T^{N+1})$ with only \emph{N} pulses
\cite{UDDexact,UDDprl,key-12}. In a UDD sequence, the $j$th control
pulse is applied at the time
\begin{eqnarray} \label{Tj} T_{j} & =T \sin^{2}(\frac{j\pi}{2N+2}),\
j=1, 2\cdots ,N.
\end{eqnarray} In most cases UDD outperforms all other
known DD control sequences, a fact already confirmed in two
beautiful experiments \cite{UDDvsCPMG,detailpra,Du-Liu}. As a dramatic
development in theory, Yang and Liu proved that UDD is universal for
suppressing single-qubit decoherence \cite{univUDD}. That is, for a
single qubit coupled with an arbitrary bath, UDD works regardless of
how the qubit is coupled to its bath.

Given the universality of UDD for suppression of single-qubit
decoherence, it becomes urgent to examine whether UDD is useful for
preserving quantum coherence of two-qubit states. This extension is
necessary and important because many quantum operations involve at
least two qubits.  Conceptually there is also a big difference
between single-qubit coherence  and two-qubit coherence: preserving
the latter often means the storage of quantum entanglement.
Furthermore, because quantum entanglement is a nonlocal property and
cannot be affected by local operations, preserving quantum
entanglement between two qubits by a control pulse sequence will
require the use of nonlocal control Hamiltonians.

In this work, by exploiting a central result in Yang and Liu's universality proof \cite{univUDD}
 for UDD in single-qubit systems and by adopting a generalized coherence measure for two-qubit states,
 we show that UDD pulse sequence does apply to two-qubit systems, at least for preserving one pre-determined
 type of quantum coherence.
 The associated control Hamiltonian is also explicitly constructed.  This significant extension from
  single-qubit to two-qubit systems opens up an exciting avenue of dynamical protection
   of quantum entanglement. Indeed, it is now possible to efficiently lock a two-qubit system
   on a desired entangled state, without any knowledge of the bath.  Encouraged by our results
   for two-qubit systems, we then show that in general, the coherence of an arbitrary
   $M$-level quantum system,
   which is characterized by our generalized coherence measure, can
also be preserved by UDD to the order of  $1+O(T^{N+1})$ with only
$N$ pulses, irrespective of how this system is coupled with its
environment.  Hence, in principle, an arbitrary (but known) quantum state of an
$m$-qubit system with $M=2^{m}$ levels can be locked by UDD,
provided that the required control Hamiltonian can be implemented
experimentally.  To establish an interesting connection with a
kicked multi-level system recently realized in a cold-atom
laboratory \cite{Jessen}, we also explicitly construct the UDD
control Hamiltonian for decoherence suppression in three-level
quantum systems.

This paper is organized as follows. In Sec. II, we
first briefly outline an important result proved by Yang and Liu \cite{univUDD};
we then present our theory for UDD in two-qubit systems, followed by an extension to multi-level quantum
systems.
In Sec. III, we present supporting results from some simple numerical experiments.  Section
 IV discusses the implications of our results and then concludes this paper.

\section{UDD Theory for two-qubit and general multi-level systems }
\subsection{On Yang-Liu's Universality Proof for Single-Qubit Systems}
For our later use we first briefly describe one central result in
Yang and Liu's work \cite{univUDD} for proving the universality of
the UDD control sequence applied to single-qubit systems. Let $C$ and $Z$ be two time-independent
Hermitian operators.  Define two unitary operator $U^{(N)}_{\pm}$ as
follows:
\begin{eqnarray}
\label{oldun}
 U^{(N)}_{\pm}(T)& =& e^{-i[C\pm (-1)^N Z](T-T_N)} \nonumber \\
&&\times\   e^{-i[C\pm (-1)^{(N-1)} Z](T_N-T_{N-1})} \cdots  \nonumber \\
           &&\times\ e^{-i[C\mp Z](T_2-T_1)} e^{-i[C\pm Z]T_1}
           .
% %=(\sigma)^{N}e^{-i\int_{T_{N}}^{T}[H_{0}+(-1)^{N}H']\mathrm{d}t}...e^{-i\int_{T_{1}}^{T_{2}}[H_{0}-H']\mathrm{d}t}e^{-i\int_{0}^{T_{1}}[H_{0}+H']\mathrm{d}t}\\
%  =e^{-iH_{0}T}(\sigma)^{N}\mathfrak{J}e^{-i\int_{0}^{T}F_{N}(t')H'_{I}(t')\mathrm{d}t'},\mathrm{\ in\ which}
\end{eqnarray}
Yang and Liu proved that for $T_j$ satisfying Eq. (\ref{Tj}), we must have
\begin{equation}
\left(U^{(N)}_{-}\right)^{\dagger} U^{(N)}_{+} = 1+O(T^{N+1}),
\label{Tn1}
\end{equation}
i.e., the product of $\left(U^{(N)}_{-}\right)^{\dagger}$ and
$U^{(N)}_{+}$ differs from unity only by the order of $O(T^{N+1})$
for sufficiently small $T$. In the interaction representation,
\begin{eqnarray}
Z_{I}(t) & \equiv & e^{iCt}Ze^{-iCt} \nonumber \\
&=& \sum_{p=0}^{\infty}\frac{(it)^{p}}{p!}\underbrace{[C,[C,...[C,Z]]]}_{p\ \mathrm{folds}},
\end{eqnarray} hence the above expression for $U^{(N)}_{\pm}$ can be rewritten in the following compact form
\begin{eqnarray}
\label{upm}
U^{(N)}_{\pm}(T)=e^{-iCT}\mathfrak{J}\left[ e^{-i\int_{0}^{T}\pm F_{N}(t)Z_{I}(t)\mathrm{d}t}\right],
\end{eqnarray}
where $T$ is the final time, $\mathfrak{J}$ denotes the time-ordering operator, and
\begin{eqnarray}
 F_{N}(t)=(-1)^j,\ {\rm for}\ \ t\in (T_j, T_{j+1}).
 \label{fn}
 \end{eqnarray}
As an important observation,  we note that though Ref. \cite{univUDD} focused on
 single-qubit decoherence in a bath,
  Eq. (\ref{Tn1}) was proved therein for arbitrary Hermitian operators $C$ and $Z$.
  This motivated us to investigate under what conditions the unitary evolution operator of
  a controlled two-qubit system plus a bath
 can assume the same form as Eq. (\ref{oldun}).

\subsection{Decoherence Suppression in Two-qubit Systems}
Quantum coherence is often characterized by the magnitude of the
off-diagonal matrix elements of the system density operator
after tracing over the bath.  In single-qubit cases, the transverse
polarization then measures the coherence and the longitudinal
polarization
 measures the population difference.
 Such a perspective is often helpful so long as
 its representation-dependent nature is well understood. In two-qubit systems or
 general multi-level systems, the concept of quantum coherence
 becomes more ambiguous because there are many off-diagonal matrix elements of the system density operator.
 Clearly then, to have a general and convenient coherence measure will be important
 for extending decoherence suppression studies beyond single-qubit systems.

Here we define a generalized polarization operator to characterize a
certain type of coherence. Specifically, associated with an
arbitrary pure state $|\Psi\rangle$ of our quantum system, we define
the following polarization operator,
\begin{eqnarray}
{\cal P}_{|\Psi \rangle} \equiv 2|\Psi\rangle\langle \Psi|-I,
\end{eqnarray}
where $I$ is the identity operator.   This polarization operator has
the following properties:
\begin{eqnarray}
\label{Y1property}
{\cal P}_{|\Psi\rangle}^2&=&I, \nonumber \\
{\cal P}_{|\Psi\rangle}|\Psi\rangle&=&|\Psi\rangle, \nonumber \\
{\cal P}_{|\Psi\rangle}|\Psi^{\perp}\rangle &=& -|\Psi^{\perp}\rangle,
\end{eqnarray}
where $|\Psi^{\perp}\rangle$ represents all other possible states of
the system that are orthogonal to $|\Psi\rangle$. Hence, if the
expectation value of ${\cal P}_{|\Psi\rangle}$ is unity, then the system
must be on the state $|\Psi\rangle$. In this sense, the expectation
value of ${\cal P}_{|\Psi\rangle}$ measures how much coherence of the
$|\Psi\rangle$-type is contained in a given system.  For example, in the
single-qubit case, ${\cal P}_{|\Psi\rangle}$ measures the longitudinal
coherence if $|\Psi\rangle$ is chosen as the spin-up state, but
measures the transverse coherence along a certain direction if
$|\Psi\rangle$ is chosen as a superposition of spin-up and spin-down
states.  Most important of all, as seen in the following, the
generalized polarization operator ${\cal P}_{|\Psi\rangle}$ can directly give the
required control Hamiltonian in order to preserve the quantum
coherence thus defined.

We now consider a two-qubit system interacting with an arbitrary bath whose self-Hamiltonian
is given by $H_{E}=c_0$.  The qubits interact with the environment via the
interaction Hamiltonian $H_{jE}=\sigma_{x}^{j}c_{x,j}+\sigma_{y}^{j}c_{y,j}+\sigma_{z}^{j}c_{z,j}$
for $j=1,2$, where $\sigma_x^{j}$, $\sigma_{y}^{j}$, and $\sigma_z^{j}$ are the standard Pauli matrices, and $c_{\alpha,j}$ are bath operators.
We further assume that the qubit-qubit interaction is given by
$H_{12}=\sum_{k,l=\{x,y,z\}}c_{kl}\sigma_{k}^{1}\sigma_{l}^{2}$, where the coefficients $c_{kl}$ may also
 depend on arbitrary bath operators.
A general total Hamiltonian describing a two-qubit system in a bath hence becomes
\begin{eqnarray}
H & = & H_{E}+H_{1E}+H_{2E}+H_{12} \nonumber \\
 & = & c_0+\sigma_{x}^{1}c_{x,1}+\sigma_{y}^{1}c_{y,1}+\sigma_{z}^{1}c_{z,1}
 + \sigma_{x}^{2}c_{x,2} \nonumber \\
 && +\ \sigma_{y}^{2}c_{y,2}+\sigma_{z}^{2}c_{z,2} +\sigma_{x}^{1}\sigma_{x}^{2}c_{xx}+\sigma_{x}^{1}\sigma_{y}^{2}c_{xy}\nonumber \\
& & +\ \sigma_{x}^{1}\sigma_{z}^{2}c_{xz}+\sigma_{y}^{1}\sigma_{x}^{2}c_{yx}+\sigma_{y}^{1}\sigma_{y}^{2}c_{yy}
 +\sigma_{y}^{1}\sigma_{z}^{2}c_{yz}\nonumber\\
 && +\ \sigma_{z}^{1}\sigma_{x}^{2}c_{zx}+\sigma_{z}^{1}\sigma_{y}^{2}c_{zy}
 +\sigma_{z}^{1}\sigma_{z}^{2}c_{zz}.
\label{eq:2 Qubits with bath-total Hamiltonian}\end{eqnarray}
For convenience each term in the above total Hamiltonian is assumed to be time independent (this assumption will be lifted in the end).

Focusing on the two-qubit subspace, the above total Hamiltonian is seen
to consist of 16 linearly-independent terms that span a natural set
of basis operators for all possible Hermitian operators acting on
the two-qubit system. This set of basis operators can be summarized
as
\begin{eqnarray}
\{X_{i}\}_{i=1,2,\cdots,16}=\{\sigma_{k}\otimes\sigma_{l}\}, \
\end{eqnarray}
where $
\sigma_{k},\sigma_{l}\in\{I,\sigma_{x},\sigma_{y},\sigma_{z}\}$,
with the orthogonality condition
$\mathrm{Trace}(X_{j}X_{k})=4\delta_{jk}$. But this choice of basis
operators is rather arbitrary. We find that this operator basis set
should be changed to new ones to facilitate operator manipulations.
In the following we examine the suppression of two types of
coherence, one is associated with non-entangled states and the other
is associated with  a Bell state.

\subsubsection{Preserving coherence associated with non-entangled states}
Let the four basis states of a two-qubit system be
$|0\rangle={|\uparrow\uparrow\rangle}$,
$|1\rangle={|\uparrow\downarrow\rangle},\
|2\rangle={|\downarrow\uparrow\rangle},\ $ and $
|3\rangle={|\downarrow\downarrow\rangle}$. The projector associated
with each of the four basis states is given by
\begin{eqnarray}
|0\rangle\langle 0| &= &P_{0}=\frac{1}{4}(1+\sigma_{z}^{1})(1+\sigma_{z}^{2}), \nonumber \\
|1\rangle\langle 1| &=& P_{1}=\frac{1}{4}(1+\sigma_{z}^{1})(1-\sigma_{z}^{2}),\nonumber \\
|2\rangle\langle 2| &= & P_{2}=\frac{1}{4}(1-\sigma_{z}^{1})(1+\sigma_{z}^{2}),\nonumber \\
|3\rangle\langle 3| & = & P_{3}=\frac{1}{4}(1-\sigma_{z}^{1})(1-\sigma_{z}^{2}).
\end{eqnarray}
As a simple example, the quantum coherence to be protected here is
assumed to be ${\cal P}_{|0\rangle}=2|0\rangle\langle 0|-I$.

We now switch to the following new set of 16 basis operators,
 \begin{eqnarray}
Y_{1} & =&  {\cal P}_{|0\rangle}= 2 P_{0}-I\nonumber \\
&= &\frac{1}{2}(-I+\sigma_{z}^{1}+\sigma_{z}^{2}+\sigma_{z}^{1}\sigma_{z}^{2}),\nonumber \\
Y_{2} & =& P_{0}+P_{1}=\frac{1}{2}(I+\sigma_{z}^{1}),\nonumber \\
Y_{3} & =& P_{0}-P_{1}+2P_{2}=\frac{1}{2}(I-\sigma_{z}^{1}+2\sigma_{z}^{2}),\nonumber\\
Y_{4} & =&P_{0}-P_{1}-P_{2}+3P_{3}\nonumber \\
&=&\frac{1}{2}(I-\sigma_{z}^{1}-\sigma_{z}^{2}+3\sigma_{z}^{1}\sigma_{z}^{2}),\nonumber\\
Y_{5} & =&|1\rangle\langle3|+|3\rangle\langle1|=\frac{1}{2}(\sigma_{x}^{1}-\sigma_{x}^{1}\sigma_{z}^{2}), \nonumber\\
Y_{6} & =&-i|1\rangle\langle3|+i|3\rangle\langle1|=\frac{1}{2}(\sigma_{y}^{1}-\sigma_{y}^{1}\sigma_{z}^{2}),\nonumber\\
Y_{7} & =&|2\rangle\langle3|+|3\rangle\langle2|=\frac{1}{2}(\sigma_{x}^{2}-\sigma_{z}^{1}\sigma_{x}^{2}),\nonumber\\
Y_{8} & =&-i|2\rangle\langle3|+i|3\rangle\langle2|=\frac{1}{2}(\sigma_{y}^{2}-\sigma_{z}^{1}\sigma_{y}^{2}),\nonumber\\
Y_{9} & =&|1\rangle\langle2|+|2\rangle\langle1|=\frac{1}{2}(\sigma_{x}^{1}\sigma_{x}^{2}+\sigma_{y}^{1}\sigma_{y}^{2}),\nonumber\\
Y_{10} & =&-i|1\rangle\langle2|+i|2\rangle\langle1|=\frac{1}{2}(\sigma_{y}^{1}\sigma_{x}^{2}-\sigma_{x}^{1}\sigma_{y}^{2}),\nonumber\\
Y_{11} & =&|0\rangle\langle1|+|1\rangle\langle0|=\frac{1}{2}(\sigma_{x}^{2}+\sigma_{z}^{1}\sigma_{x}^{2}),\nonumber\\
Y_{12} & =&-i|0\rangle\langle1|+i|1\rangle\langle0|=\frac{1}{2}(\sigma_{y}^{2}+\sigma_{z}^{1}\sigma_{y}^{2}),\nonumber\\
Y_{13} & =&|0\rangle\langle2|+|2\rangle\langle0|=\frac{1}{2}(\sigma_{x}^{1}+\sigma_{x}^{1}\sigma_{z}^{2}),\nonumber\\
Y_{14} & =&-i|0\rangle\langle2|+i|2\rangle\langle0|=\frac{1}{2}(\sigma_{y}^{1}+\sigma_{y}^{1}\sigma_{z}^{2}),\nonumber\\
Y_{15} & =&|0\rangle\langle3|+|3\rangle\langle0|=\frac{1}{2}(\sigma_{x}^{1}\sigma_{x}^{2}-\sigma_{y}^{1}\sigma_{y}^{2}),\nonumber\\
Y_{16} &
=&-i|0\rangle\langle3|+i|3\rangle\langle0|=\frac{1}{2}(\sigma_{x}^{1}\sigma_{y}^{2}+\sigma_{y}^{1}\sigma_{x}^{2}).
\end{eqnarray}
%MM; previously Y_6 lack of factor 1/2

Using this new set of basis operators for a two-qubit system, the total
Hamiltonian becomes a linear combination of the $Y_{j}\ (j=1-16)$
operators defined above, i.e.,
\begin{eqnarray}
H=\sum_{j=1}^{16}W_{j}Y_{j},
\end{eqnarray} where $W_{j}$ are the
expansion coefficients that can contain arbitrary bath operators.
The above new set of basis operators have the following properties.
First, the operator $Y_1$ in this set are identical with ${\cal P}_{|0\rangle}$ and hence
also satisfies the interesting properties described by Eq. (\ref{Y1property}).
Second,
\begin{eqnarray}
[Y_{j},Y_{1}]&=&0,\ {\rm for}\ j=1,2,\cdots,10; \nonumber \\
\{Y_{j},Y_{1}\}_{+}&=&0, \ {\rm for}\ j=11,12,\cdots,16,
\end{eqnarray}
where $[\cdot]$ represents the commutator and $\{\cdot\}_{+}$
represents an anti-commutator. Third,
\begin{eqnarray}
\left[\sum_{i=1}^{10}A_{i}Y_{i},
\sum_{j=11}^{16}B_{j}Y_{j}\right]&=&\sum_{j=11}^{16}C_{j}Y_{j},
\nonumber \\ \left(\sum_{i=1}^{10}A_{i}Y_{i}\right) \left(
\sum_{j=1}^{10}B_{j}Y_{j}\right)&=&\sum_{j=1}^{10}C_{j}Y_{j},
\nonumber \\ \left(\sum_{i=11}^{16}A_{i}Y_{i}\right)
\left(\sum_{j=11}^{16}B_{j}Y_{j}\right)&=&\sum_{j=1}^{10}C_{j}Y_{j}.
\label{antic square}
\end{eqnarray}
%New Equations added
With these observations, we next split the total uncontrolled
Hamiltonian into two terms, i.e.,  $H=H_{0}+H'$, where
\begin{eqnarray}
H_{0}=W_{1}Y_{1}+W_{2}Y_{2}+\cdots +W_{10}Y_{10}, \end{eqnarray} and
\begin{eqnarray}
H'=W_{11}Y_{11}+\cdots +W_{16}Y_{16}. \end{eqnarray} Evidently, we have
the anti-commuting relation
\begin{eqnarray}
\{Y_1,H'\}_{+}=0, \label{antic}
\end{eqnarray}
an important fact for our proof below.

Consider now the following control Hamiltonian describing a sequence
of extended UDD $\pi$-pulses
\begin{eqnarray}
\label{control1}
H_{c}=\sum_{j=1}^{N}\pi \delta(t-T_j)\frac{Y_1}{2}.
\end{eqnarray}
After the $N$ control pulses, the unitary evolution operator for the
whole system of the two qubits plus a bath is given by ($\hbar=1$ throughout)
\begin{eqnarray}
{U}(T) & = & e^{-i[H_{0}+H'](T-T_N)} (-iY_1) \nonumber \\
& &\ \times\ e^{-i[H_{0}+H'](T_N-T_{N-1})} (-iY_1) \nonumber \\
& &\ \cdots \nonumber \\
 & &\ \times\  e^{-i[H_{0}+H'](T_3-T_2)} (-iY_1) \nonumber \\
& &\ \times\ e^{-i[H_{0}+H'](T_2-T_1)} (-iY_1)\nonumber \\
& &\ \times\ \ e^{-i[H_{0}+H']T_1}.
\label{Ut}
  \end{eqnarray}
We can then take advantage of the anti-commuting relation of Eq.
(\ref{antic}) to exchange the order between $(-iY_1)$ and the
exponentials in the above equation, leading to
\begin{eqnarray}
U(T) & = & (-iY_1)^{N} e^{-i[H_{0}+(-1)^{N} H'](T-T_N)}  \nonumber \\
& &\ \times\ e^{-i[H_{0}+(-1)^{N-1}H'](T_N-T_{N-1})}  \nonumber \\
& &\ \cdots \nonumber \\
 & &\ \times\  e^{-i[H_{0}+H'](T_3-T_2)}  \nonumber \\
& &\ \times\ e^{-i[H_{0}-H'](T_2-T_1)} \nonumber \\
& &\ \times\ e^{-i[H_{0}+H']T_1} \nonumber \\
&=& (-iY_1)^{N}e^{-iH_0 T} \mathfrak{J}\left[ e^{-i\int_{0}^{T}
F_{N}(t)H_{I}'(t)\mathrm{d}t}\right] \nonumber
 \\
 &\equiv & (-iY_1)^{N} {\cal U}^{(N)}_{+}(T). \label{un}
  \end{eqnarray}
Here $F_{N}(t)$ is already defined in Eq. (\ref{fn}), the second
equality is obtained by using the interaction representation, with
$H_{I}'(t)\equiv e^{iH_0 t}H_I e^{-iH_0 t}$, and the last line defines the operator
${\cal U}^{(N)}_{+}(T)$. Clearly, ${\cal U}^{(N)}_{+}$ is exactly in the form of ${U}^{(N)}_{+}$
defined in Eqs. (\ref{oldun}) and (\ref{upm}), with $H_0$ replacing $C$ and $H'$
replacing $Z$.  This observation motivates us to define
\begin{eqnarray}
{\cal U}^{(N)}_{-}(T) \equiv  e^{-iH_0 T} \mathfrak{J}\left[ e^{-i\int_{0}^{T}-
       F_{N}(t)H_{I}'(t)\mathrm{d}t}\right],
\end{eqnarray}
which is completely in parallel with ${U}^{(N)}_{-}$ defined in Eq.
(\ref{upm}).
 As such, Eq. (\ref{Tn1}) directly leads to
\begin{eqnarray}
\left({\cal U}^{(N)}_{-}\right)^{\dagger} {\cal U}^{(N)}_{+} = 1+O(T^{N+1}).
\label{ourTn1}
\end{eqnarray}
%MM

With  Eq. (\ref{ourTn1}) obtained we can now evaluate the coherence
measure. In particular, for an arbitrary initial state given by the
density operator $\rho_{i}$, the expectation value of  ${\cal
P}_{|0\rangle}$ at time $T$ is given by
\begin{eqnarray}
  && {\rm Trace} \{ U(T) \rho_{i} U^{\dagger}(T){\cal P}_{|0\rangle}\} \nonumber \\
 &=& {\rm Trace} \{ (-iY_1)^{N}{\cal U}^{(N)}_{+} \rho_{i} \left({\cal U}^{(N)}_{+}\right)^{\dagger} (iY_1)^{N} {\cal P}_{|0\rangle}\} \nonumber  \\
&=& {\rm Trace} \{ (-iY_1)^{N}{\cal U}^{(N)}_{+} \rho_{i}{\cal P}_{|0\rangle}  \left({\cal U}^{(N)}_{-}\right)^{\dagger} (iY_1)^{N}\} \nonumber  \\
&=& {\rm Trace} \{  \left({\cal U}^{(N)}_{-}\right)^{\dagger} {\cal U}^{(N)}_{+} \rho_{i}{\cal P} _{|0\rangle}\} \nonumber \\
&=& {\rm Trace} \{ \rho_{i}{\cal P}_{|0\rangle} \} \left[1+O(T^{N+1})\right],
\label{ourmain}
\end{eqnarray}
where we have used ${\cal P}_{|0\rangle}=Y_1$, $Y_1^2=I$, and the
anti-commuting relation between ${\cal P}_{|0\rangle}$ and $H'$.
 Equation (\ref{ourmain}) clearly demonstrates that, as a result of the UDD sequence of $N$ pulses,
 the expectation value of ${\cal P}_{|0\rangle}$ is preserved to the order of $1+O(T^{N+1})$,
 for an arbitrary initial state.  If the initial state is set to be $|0\rangle$, i.e., ${\rm Trace} \{ \rho_{i}{\cal P}_{|0\rangle} \}=1$,
 then the expectation value of ${\cal P}_{|0\rangle}$ remains to be $1+O(T^{N+1})$ at time $T$,
 indicating that the UDD sequence has locked the system on the state $|0\rangle={|\uparrow\uparrow\rangle}$.

%Similar to Yang and Liu's work \cite{univUDD}, the unitary propagator ${\cal U}^{(N)}_{+}$ can be written as
%\begin{eqnarray}
%{\cal U}^{(N)}_{+}&=&{e}^{-i{H}_{\mathrm{eff}}T+O({T}^{N+1})} \label{spin polar up-up}
%\end{eqnarray}
%where ${H}_{\mathrm{eff}}$ commutes with ${Y}_{1}$ because the even-order terms from the expansion of the time-ordering operator in Eq. (\ref{un}) %contains only  ${Y}^{{n}_{11}}_{11}{Y}^{{n}_{12}}_{12}{Y}^{{n}_{13}}_{13}{Y}^{{n}_{14}}_{14}{Y}^{{n}_{15}}_{15}{Y}^{{n}_{16}}_{16}$ with %${n}_{11}+{n}_{12}+{n}_{13}+{n}_{14}+{n}_{15}+{n}_{16}$ is even which all commutes with ${Y}_{1}$according to Eq. (\ref{antic square}).

In our proof of the UDD applicability in preserving the coherence
${\cal P}_{|\Psi\rangle}$ associated with a non-entangled state, the first
important step is to construct the control operator $Y_1={\cal P}_{|\Psi\rangle}$ and then
the control Hamiltonian $H_c$. As is clear from Eq.
(\ref{Y1property}), each application of the control operator
$Y_1={\cal P}_{|0\rangle}$ leaves the state $|0\rangle$ intact but
induces a negative sign for all other two-qubit states.  It is interesting to compare the control operator
$Y_1$ with what can be intuitively expected from early single-qubit UDD results.
Suppose that the two qubits are unrelated at all, then in order to suppress the spin flipping of the first qubit (second qubit),
we need a control operator $\sigma_z^1$ ($\sigma_z^2$). As such,  an intuitive single-qubit-based control Hamiltonian would be
\begin{eqnarray}
  H_{c,\text{single}}=\frac{\pi}{2}\sum_{j=1}^{N}\delta(t-T_j)(\sigma_{z}^{1}+\sigma_{z}^{2}).
  \label{intu}
  \end{eqnarray}
This intuitive control Hamiltonian differs from  Eq.
(\ref{control1}), hinting an importance difference between two-qubit
and single-qubit cases.  Indeed, here the qubit-qubit interaction or
the system-environment coupling may directly cause a double-flipping
error ${|\uparrow\uparrow\rangle} \rightarrow
{|\downarrow\downarrow\rangle}$, which cannot be suppressed by
$H_{c,\text{single}}$.  The second key step is to split the
Hamiltonian $H$ into two parts $H_0$ and $H'$, with the former
commuting with $Y_1$ and the latter anti-commuting with $Y_1$. Once
these two steps are achieved, the remaining part of our proof
becomes straightforward by exploiting Eq. (\ref{ourTn1}). These
understandings suggest that it should be equally possible to
preserve the coherence associated with entangled two-qubit states.

\subsubsection{Preserving coherence associated with entangled states}
Consider a different coherence property as defined by our generalized polarization operator ${\cal P}_{|\Psi\rangle}$, with $|\Psi\rangle$
taken as
a Bell state
\begin{eqnarray}
\label{bell}
|\tilde{0}\rangle=\frac{1}{\sqrt{2}}[{|\uparrow\downarrow\rangle}+{|\downarrow\uparrow\rangle}].
\end{eqnarray}
The other three orthogonal basis states for the two-qubit system are
now denoted as $|\tilde{1}\rangle$, $|\tilde{2}\rangle$,
$|\tilde{3}\rangle$. For example, they can be assumed to be
${|\tilde{1}\rangle} = \frac{1}{\sqrt{2}}[|\uparrow\uparrow\rangle +
|\downarrow\downarrow\rangle]$,    ${|\tilde{2}\rangle} =
\frac{1}{\sqrt{2}}[{|\uparrow\uparrow\rangle} -
{|\downarrow\downarrow\rangle}]$, and ${|\tilde{3}\rangle} =
\frac{1}{\sqrt{2}}[{|\uparrow\downarrow\rangle} -
{|\downarrow\uparrow\rangle}]$. To preserve such a new type of
coherence, we follow our early procedure to first construct a
control operator $\tilde{Y}_1$ and then a new set of basis
operators. In particular, we require
\begin{eqnarray}
\tilde{Y}_1&=& {\cal P}_{|\tilde{0}\rangle} = 2|\tilde{0}\rangle\langle \tilde{0}|- I \nonumber \\
&=& \frac{1}{2}(-I+\sigma_{x}^{1}\sigma_{x}^{2}+\sigma_{y}^{1}\sigma_{y}^{2}-\sigma_{z}^{1}\sigma_{z}^{2}).
\end{eqnarray}
We then construct other 9 basis operators that all commute with $\tilde{Y}_1$, e.g.,
\begin{eqnarray}
\tilde{Y}_{2} & =& \frac{1}{2}(I+\sigma_{x}^{1}\sigma_{x}^{2}),\nonumber \\
\tilde{Y}_{3} & =& \frac{1}{2}(I-\sigma_{x}^{1}\sigma_{x}^{2}+2\sigma_{y}^{1}\sigma_{y}^{2}),\nonumber \\
\tilde{Y}_{4} & =& \frac{1}{2}(I-\sigma_{x}^{1}\sigma_{x}^{2}-\sigma_{y}^{1}\sigma_{y}^{2}-3\sigma_{z}^{1}\sigma_{z}^{2}),\nonumber \\
\tilde{Y}_{5} & =& \frac{1}{2}(\sigma_{z}^{1}\sigma_{x}^{2}-\sigma_{x}^{1}\sigma_{z}^{2}),\nonumber \\
\tilde{Y}_{6} & =& \frac{1}{2}(\sigma_{y}^{2}-\sigma_{y}^{1}),\nonumber \\
\tilde{Y}_{7} & =& \frac{1}{2}(\sigma_{x}^{2}-\sigma_{x}^{1}),\nonumber \\
\tilde{Y}_{8} & =& -\frac{1}{2}(\sigma_{y}^{1}\sigma_{z}^{2}-\sigma_{z}^{1}\sigma_{y}^{2}),\nonumber \\
\tilde{Y}_{9} & =&\frac{1}{2}(\sigma_{z}^{1}+\sigma_{z}^{2}),\nonumber \\
\tilde{Y}_{10} & =& -\frac{1}{2}(\sigma_{x}^{1}\sigma_{y}^{2}+\sigma_{y}^{1}\sigma_{x}^{2}).
\end{eqnarray}
The remaining 6 linearly independent basis operators are found to be anti-commuting with $\tilde{Y}_1$.
They can be written as \begin{eqnarray}
\tilde{Y}_{11} & =&\frac{1}{2}(\sigma_{x}^{1}+\sigma_{x}^{2}),\nonumber \\
\tilde{Y}_{12} & =&-\frac{1}{2}(\sigma_{y}^{1}\sigma_{z}^{2}+\sigma_{z}^{1}\sigma_{y}^{2}),\nonumber \\
\tilde{Y}_{13} & =&\frac{1}{2}(\sigma_{x}^{1}\sigma_{z}^{2}+\sigma_{z}^{1}\sigma_{x}^{2}),\nonumber \\
\tilde{Y}_{14} & =&-\frac{1}{2}(\sigma_{y}^{1}+\sigma_{y}^{2}),\nonumber \\
\tilde{Y}_{15} & =&\frac{1}{2}(\sigma_{z}^{1}-\sigma_{z}^{2}),\nonumber \\
\tilde{Y}_{16} & =&\frac{1}{2}(\sigma_{x}^{1}\sigma_{y}^{2}-\sigma_{y}^{1}\sigma_{x}^{2}).
\end{eqnarray}
The total Hamiltonian can now be rewritten as
$H=\tilde{H}_{0}+\tilde{H}'$, in which
\begin{eqnarray}
\tilde{H}_{0}=\tilde{W}_{1}\tilde{Y}_{1}+\tilde{W}_{2}\tilde{Y}_{2}+\cdots +\tilde{W}_{10}\tilde{Y}_{10} \end{eqnarray} and
\begin{eqnarray}
\tilde{H}'=\tilde{W}_{11}\tilde{Y}_{11}+\cdots +\tilde{W}_{16}\tilde{Y}_{16}. \end{eqnarray}

It is then evident that if we apply the following control Hamiltonian, i.e.,
\begin{eqnarray}
\label{control2}
\tilde{H}_{c}&=&\sum_{j=1}^{N}\pi \delta(t-T_j)\frac{\tilde{Y}_1}{2} \nonumber \\
&=& \sum_{j=1}^{N}\frac{\pi}{4}
\delta(t-T_j)(-I+\sigma_{x}^{1}\sigma_{x}^{2}+\sigma_{y}^{1}\sigma_{y}^{2}-\sigma_{z}^{1}\sigma_{z}^{2}),
\end{eqnarray}
the time evolution operator of the controlled total system becomes
entirely parallel to Eqs. (\ref{Ut}) and (\ref{un}) (with an
arbitrary operator $O$ replaced by $\tilde{O}$).  Hence,
using the $N$ control pulse described by Eq. (\ref{control2}),
the quantum coherence defined by the expectation value of
${\cal P}_{|\tilde{0}\rangle}$ can be preserved up to $1+O(T^{N+1})$, for
an arbitrary initial state. If the initial state is already the Bell state
$|\tilde{0}\rangle$ (i.e., coincides with the $|\Psi\rangle$ that defines our coherence measure ${\cal P}_{|\Psi\rangle}$),
 then our UDD control sequence locks the system
on this Bell state with a fidelity $1+O(T^{N+1})$, no matter how the
system is coupled to its environment.

The constant term in the control Hamiltonian $\tilde{H}_c$ can be dropped because it only induces an overall phase of the evolving state.
All other terms in $\tilde{H}_c$ represent two-body and hence nonlocal control.  This confirms our
initial expectation that suppressing the decoherence
of entangled two-qubit states is more involving than in single-qubit cases.

We have also considered the preservation of another Bell state
$\frac{1}{\sqrt{2}}[{|\uparrow\downarrow\rangle}-{|\downarrow\uparrow\rangle}]$.
Following the same procedure outlined above, one finds that the
required UDD control Hamiltonian should be given by
 \begin{eqnarray}
\tilde{H}_{c}&=& -\sum_{j=1}^{N}\frac{\pi}{4} \delta(t-T_j)(I+\sigma_{x}^{1}\sigma_{x}^{2}+\sigma_{y}^{1}\sigma_{y}^{2}+\sigma_{z}^{1}\sigma_{z}^{2}),
\label{enc2}
\end{eqnarray}
which is a pulsed Heisenberg interaction Hamiltonian. Such an isotropic control Hamiltonian is consistent with the fact the singlet Bell state defining our quantum
coherence measure is also isotropic.

\subsection{UDD in M-level systems}
Our early consideration for two-qubit systems suggests a general strategy for establishing UDD in
an arbitrary $M$-level system.  Let $|0\rangle, |1\rangle, \cdots, |M-1\rangle$ be
the $M$ orthogonal basis states for an $M$-level system. Their associated projectors are
defined as $P_{j}\equiv |j\rangle\langle j|$, with $j=0, 1, \cdots, M-1$.  Without loss of generality
we consider the quantum coherence to be preserved is of the $|0\rangle$-type, as characterized by
${\cal P}_{|0\rangle}=2|0\rangle\langle 0|-I$.
As learned from Sec. II-B, the important control operator is then
\begin{eqnarray}
V_1 & =& {\cal P}_{|0\rangle}= 2P_{0}-I,
\end{eqnarray}
with $V_1^2=I$. A UDD sequence of this control operator can be
achieved by the following control Hamiltonian
\begin{eqnarray}
\label{Mc}
\tilde{H}_{c}&=& \sum_{j=1}^{N}\pi \delta(t-T_j)\frac{V_1}{2}.
\end{eqnarray}

In the $M$-dimensional Hilbert space, there are totally $M^2$ linearly independent Hermitian operators.  We now divide the $M^2$ operators
into two groups, one commutes with $V_1$ and the other anti-commutes with $V_1$.  Specifically, the following $M-1$ operators
\begin{eqnarray}
V_{2} & =& P_{0}+P_{1},\nonumber \\
V_3 &= & P_0 -P_1 +2P_2, \nonumber \\
&\cdots &\nonumber \\
V_{M} & =& P_{0}-P_{1}-...-P_{M-2}+(M-1)P_{M-1}
\end{eqnarray}
evidently commutes with $V_1$. In addition, other $(M-2)(M-1)$ basis operators,
denoted $V_{M+1},V_{M+2},\cdots,V_{M+(M-2)(M-1)}$,  also commute with $V_1$.  This is the case because we can construct the following $\frac{1}{2}(M-2)(M-1)$ basis operators
\begin{eqnarray}
|k\rangle\langle l |+| l\rangle\langle k|
\end{eqnarray} with $0<k<M$ and $k<l<M$.
The other $\frac{1}{2}(M-2)(M-1)$ basis operators that commute with
$V_1$ are constructed as
\begin{eqnarray}
-i|k\rangle\langle l |+i| l\rangle\langle k|,
\end{eqnarray} also with $0<k<M$ and $k<l<M$.
All the remaining $2(M-1)$ basis operators are found to anti-commute with ${V}_{1}$. Specifically, they can be written as
\begin{eqnarray}
V_{M+(M-1)(M-2)+2l-1} & =& |0\rangle\langle l|+|l\rangle\langle0|; \nonumber \\
V_{M+(M-1)(M-2)+2l} & =& -i|0\rangle\langle l|+i|l\rangle\langle0|,
\end{eqnarray}
where $1\leq l\leq M-1$.

The total Hamiltonian for an uncontrolled $M$-level system interacting with a bath can now be written as
\begin{eqnarray}
H_{M}&=& H_{0}+H', \nonumber \\
 H_{0}&=&\sum_{j=1}^{M^{2}-2M+2}W_{j}V_{j}, \nonumber \\
H'&=&\sum_{j=M^{2}-2M+3}^{M^{2}}W_{j}V_{j},
\end{eqnarray}
where $W_{j}$ are the expansion coefficients that may contain arbitrary bath operators.

With the UDD control sequence described in Eq. (\ref{Mc}) tuned on,
the unitary evolution operator can be easily investigated using
$[V_1,H_{0}]=0$ and $\{V_1,H'\}_{+}=0$. Indeed, it takes exactly the
same form (with $Y_1\rightarrow V_1$) as in Eq. (\ref{un}). We can
then conclude that, the quantum coherence property
${\cal P}_{|\Psi\rangle}$ associated with an arbitrarily pre-selected state
$|\Psi\rangle$ in an $M$-level system can be preserved with a
fidelity $1+O(T^{N+1})$, with only $N$ pulses. For an $m$-qubit
system, $M=2^m$. In such a multi-qubit case, our result here indicates the following:
if the initial state of an $m$-qubit system is known, then by
(i) setting $|\Psi\rangle$ the same as this initial state, and then (ii) setting ${\cal P}_{|\Psi\rangle}$ as the control operator,
the known initial state will be efficiently locked by UDD.  Certainly, realizing the required control Hamiltonian for a multi-qubit system
may be experimentally challenging.

Recently, a  multi-level system subject to pulsed external fields is experimentally realized in a cold-atom laboratory \cite{Jessen}.
To motivate possible experiments of UDD using an analogous setup, in the following we consider the case of $M=3$ in detail.  To gain more insights into the control operator $V_1$, here we
use angular momentum operators in the $J=1$ subspace to express all the nine basis operators.
Specifically, using the eigenstates of the $j_z$ operator as our representation,  we have
\begin{eqnarray}
j_{x}& =& \frac{1}{\sqrt{2}}\left(\begin{array}{ccc}
 & 1\\
1 &  & 1\\
 & 1\end{array}\right), \nonumber \\
 j_{y}&=&\frac{1}{\sqrt{2}}\left(\begin{array}{ccc}
 & -i\\
i &  & -i\\
 & i\end{array}\right), \nonumber \\
   j_{z}& =&\left(\begin{array}{ccc}
1\\
 & 0\\
 &  & -1\end{array}\right).\end{eqnarray}
As an example, we use the state $(1,0,0)^{\text{T}}$ to define our coherence measure.
The associated control operator $V_1$ is then found to be
\begin{eqnarray}
\label{control3}
V_1=j_{z}+j_{z}^{2}-I.
\end{eqnarray}
Interestingly, this control operator involves a nonlinear function of
the angular momentum operator $j_z$.  This requirement can be experimentally fulfilled, because
realizing such kind of operators in a pulsed fashion is one main
achievement of Ref. \cite{Jessen}, where a ``kicked-top" system is
realized for the first time. The two different contexts, i.e., UDD
by instantaneous pulses and the delta-kicked top model for
understanding quantum-classical correspondence and quantum chaos
 \cite{Jessen,haake,Gongprl}, can thus be connected to each other.

For the sake of completeness, we also present below those operators that commute with $V_1$, namely,
\begin{eqnarray}
V_{2} & = &I+\frac{1}{2}j_{z}-\frac{1}{2}j_{z}^{2}, \nonumber\\
V_{3} & =& -I-\frac{1}{2}j_{z}+\frac{5}{2}j_{z}^{2}, \nonumber\\
V_{4} & =& -\frac{1}{\sqrt{2}}(j_{+}j_{z}+j_{z}j_{-}), \nonumber \\
V_5 & =& \frac{i}{\sqrt{2}}(j_{+}j_{z}-j_{z}j_{-}),
\end{eqnarray}
where $j_{\pm}=j_{x}\pm ij_{y}$; and those operators that anti-commute with $V_1$, namely,
\begin{eqnarray}
V_{6} & =& \frac{1}{\sqrt{2}}(j_{z}j_{+}+j_{-}j_{z}), \nonumber \\
V_{7} & =&\frac{i}{\sqrt{2}}(j_{-}j_{z}-j_{z}j_{+}),\nonumber \\
V_{8} & =&\frac{1}{2}(j_{+}^{2}+j_{-}^{2}), \nonumber \\
V_{9} & =& \frac{i}{2}(j_{-}^{2}-j_{+}^{2}).
\end{eqnarray}
%previously V_8 and V_9 lack of factor 1/2
Some linear combinations of these operators will be required to
construct the control Hamiltonian to preserve the coherence
associated with other states.

\section{Simple Numerical Experiments}

\begin{figure}[t]
\begin{center}
%\vspace*{-0.5cm}
\par
\resizebox*{10cm}{8cm}{\includegraphics*{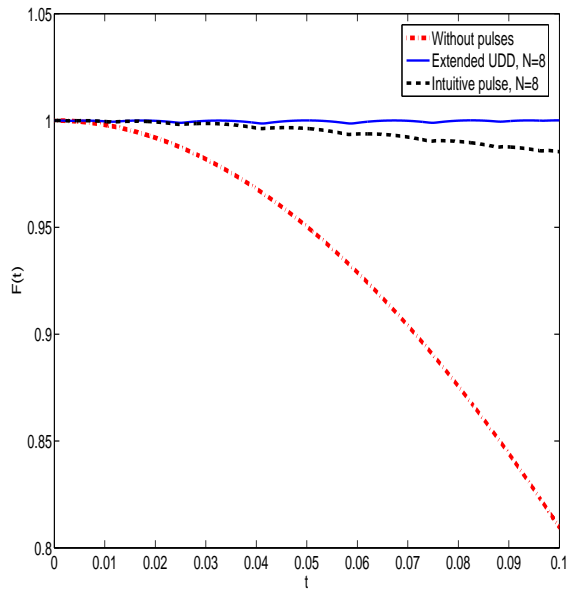}}
\end{center}
\par
 \caption{(color online) Expectation value of the coherence measure ${\cal P}_{|\Psi\rangle}$,
 denoted $F(t)$, as a function of time in dimensionless units, with $|\Psi\rangle$
 being the
 non-entangled state ${|\uparrow\uparrow\rangle}$ of a two-qubit system. The bath responsible for the decoherence
 is modeled by a three-spin system detailed in the text. The bottom curve is without any control and the decoherence is significant.
 The middle curve is calculated from a control Hamiltonian intuitively based on two
 independent qubits.
The top solid curve represents significant decoherence suppression
due to our two-qubit UDD control Hamiltonian described by Eq.
(\ref{control1}).}
\end{figure}

To further confirm the UDD control sequences we explicitly constructed above,  we have performed some simple numerical experiments.
We first consider
a model of a two-spin system coupled to a bath of three spins. The total
Hamiltonian in dimensionless units is hence given by \begin{eqnarray}
H & = & \sum_{m=3}^{5}\sum_{j=\{x,y,z\}}b_{j,m}\sigma_{j}^{m}\nonumber \\
 &  & + \sum_{n=1}^{5}\sum_{k=\{x,y,z\}}\sum_{m>n}^{5}\sum_{j=\{x,y,z\}}c_{jk}\sigma_{j}^{m}\sigma_{k}^{n}\nonumber \\
 &  & +\ H_{c},\end{eqnarray}
where the first two spins constitute the two-qubit system in the absence of
any external field, $H_{c}$ represents the UDD control Hamiltonian,
and the coefficients $b_{j,m}$ and $c_{jk}$ take randomly chosen
values in $[0,1]$ in dimensionless units. In addition, to be more
realistic, we replace the instantaneous $\delta(t-T_{j})$ function
in our control Hamiltonians by a Gaussian pulse, i.e.,
$(1/c\sqrt{\pi})e^{-[(t-T_{j})^{2}/c^{2}]}$, with $c=T/100$ unless
specified otherwise.  Further, we set $T=0.1$, because this scale is
comparable to the decoherence time scale.

\begin{figure}[t]
\begin{center}
%\vspace*{-0.5cm}
\par
\resizebox*{10cm}{8cm}{\includegraphics*{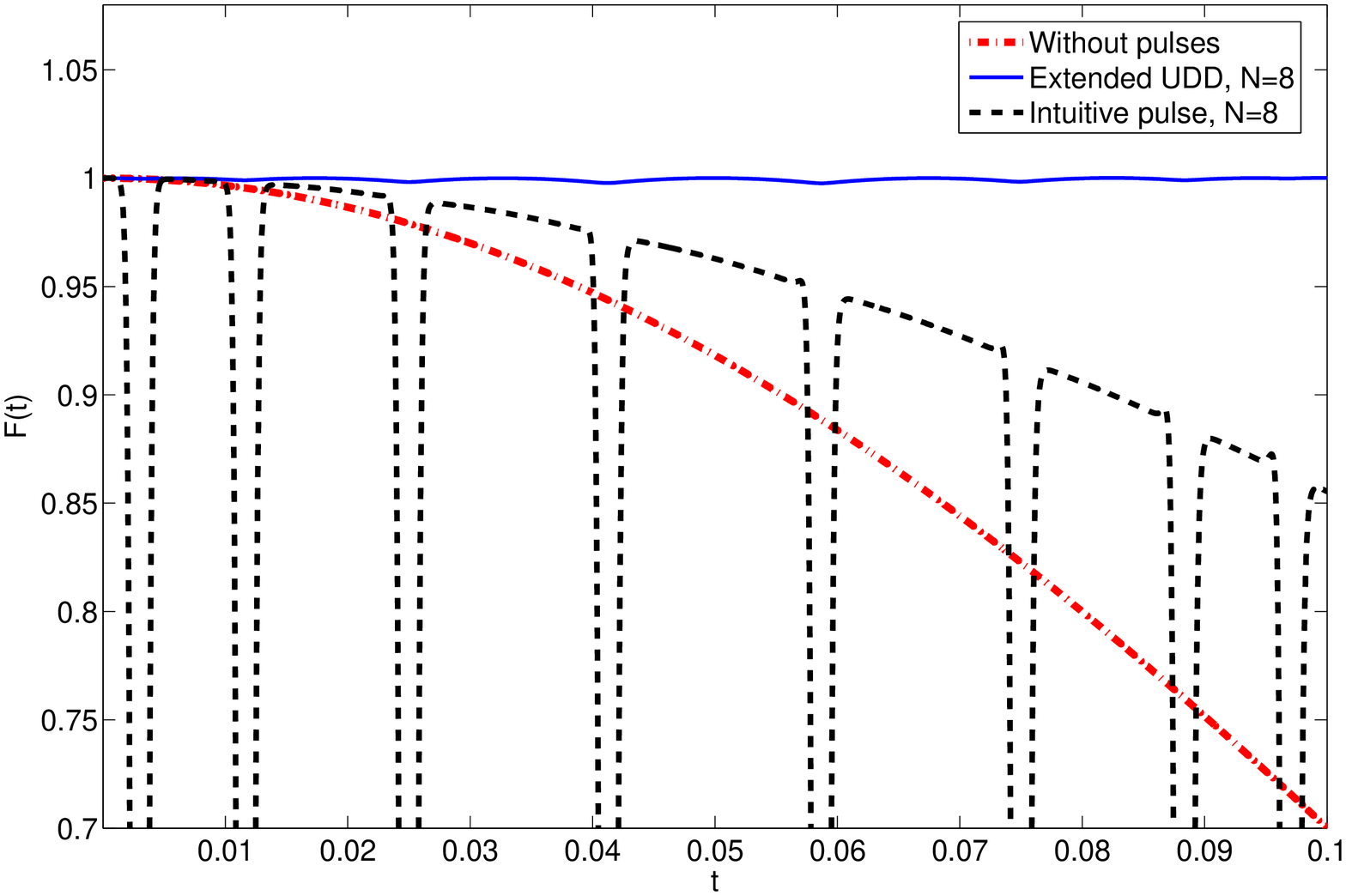}}
\end{center}
\par
 \caption{(color online) Same as in Fig. 1, but for ${\cal P}_{|\Psi\rangle}$ associated with a Bell state defined in Eq. (\ref{bell}).
  The smooth dashed curve represents significant decoherence without control. The drastically oscillating
  dashed curve is calculated from an intuitive single-qubit-based
  control Hamiltonian, showing strong population transfer from the initial state to other two-qubit states.
  The top solid curve represents signficant decoherence suppression due to
   our two-qubit UDD control sequence in Eq. (\ref{control2}).}
\end{figure}

\begin{figure}[t]
\begin{center}
%\vspace*{-0.5cm}
\par
\resizebox*{10cm}{8cm}{\includegraphics*{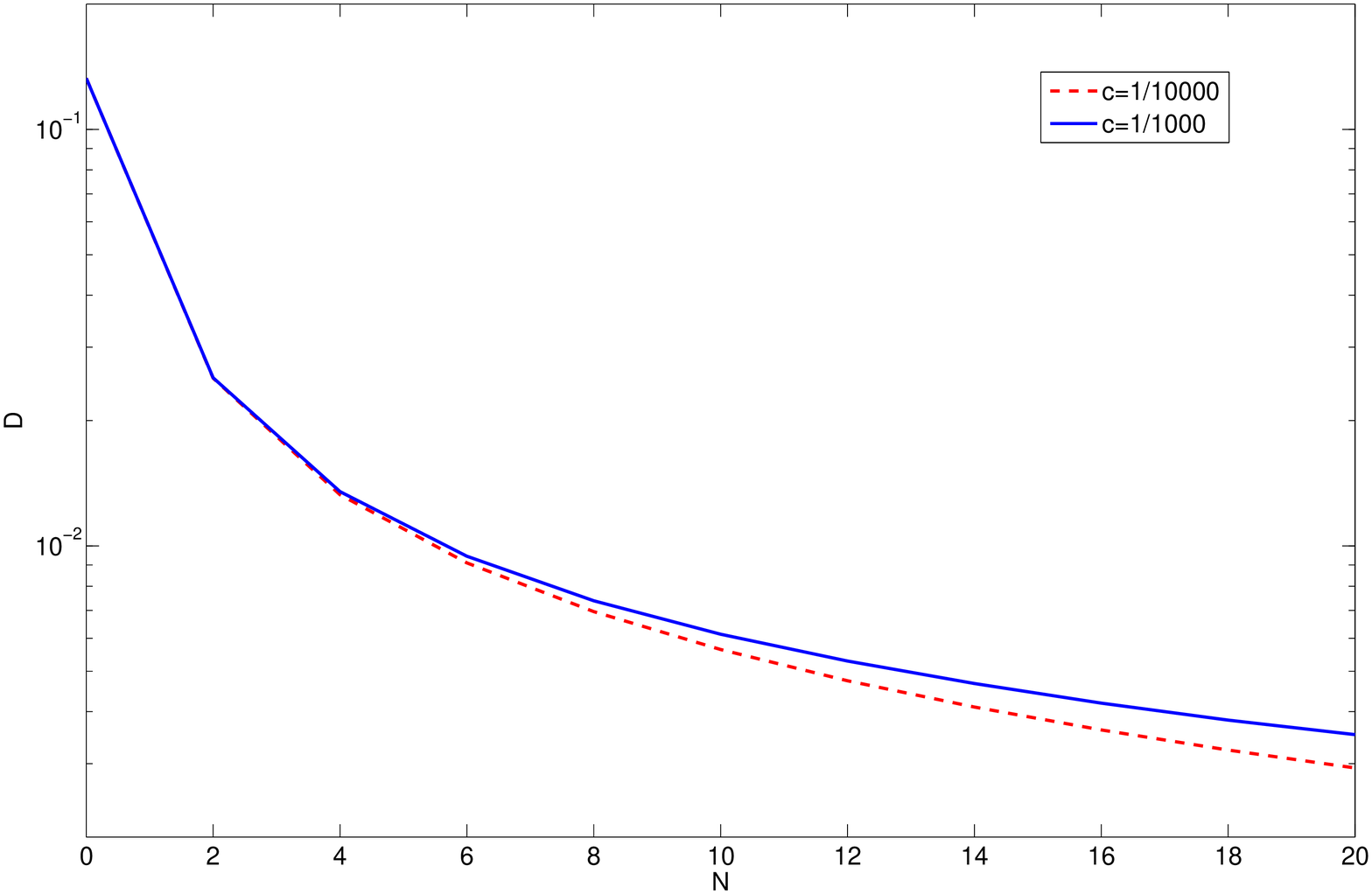}}
\end{center}
\par
 \caption{(color online)  The time-averaged distance $D$ between the actual density matrix from that
 of a completely locked Bell state, for $c=T/100$ and $c=T/1000$, versus the number of UDD pulses. The initial state is the same as in Fig. 2.}
\end{figure}

Figure 1 depicts the time dependence of the expectation value of the
coherence measure ${\cal P}_{|\Psi\rangle}$, denoted $F(t)$, with $|\Psi\rangle$
being the
 non-entangled state ${|\uparrow\uparrow\rangle}$ of the two-qubit system. The
 initial state of the system is
 also taken as the non-entangled state ${|\uparrow\uparrow\rangle}$.  As is evident from the uncontrolled case (bottom curve) ,
 the decoherence time scale without any decoherence suppression is of the order 0.1
 in dimensionless units.  Turning on the
 two-qubit UDD control sequence described by Eq. (\ref{control1}) for $N=8$,
 the decoherence (top solid curve) is seen to be greatly suppressed.
 We have also examined the decoherence suppression using a
 UDD sequence based on the single-qubit-based intuitive control Hamiltonian $H_{c,\text{single}}$ described by Eq. (\ref{intu}).
 As shown in Fig. 1,
 $H_{c,\text{single}}$ can
  only produce unsatisfactory decoherence suppression.

Similar results are obtained in Fig. 2, where we aim to preserve the coherence
measure ${\cal P}_{|\Psi\rangle}$ associated with the Bell state defined in
Eq. (\ref{bell}). Apparently, with the assistance of our two-qubit
UDD control sequence, the system is seen to be locked on the Bell
state with a fidelity close to unity at all times.  Figure 2 also
presents the parallel result if the control Hamiltonian is given by
$H_{c,\text{single}}$ shown in Eq. (\ref{intu}). The drastic oscillation
of $F(t)$ in this case indicates that strong population oscillation
occurs, thereby demonstrating again the difference between
single-qubit decoherence suppression and two-qubit decoherence
suppression.

Using the same initial state as in
Fig. 2,  Fig. 3 depicts $\overline{D} \equiv \frac{1}{2T}\int_{0}^{T}  ||\rho(t)-\rho_i|| dt $,
i.e., the time-averaged distance
between the actual time-evolving density matrix from that of a completely locked
Bell state, for $c=T/100$ and $c=T/1000$, with different number
of UDD pulses. It is seen that, at least for the number of UDD
pulses considered here,
 $c=T/100=1/1000$ (about one hundredth of the decoherence time scale)
already suffices to preserve a Bell state. That is, there seems to
be no need to use much shorter pulses such as $c=T/1000=1/10000$,
because the case of $c=T/1000$ (dashed line) in Fig. 3 shows little
improvement as compared with the case of $c=T/100$ (solid line).
This should be of practical interest for experimental studies of
two-qubit decoherence suppression.

\begin{figure}[t]
\begin{center}
%\vspace*{-0.5cm}
\par
\resizebox*{10cm}{8cm}{\includegraphics*{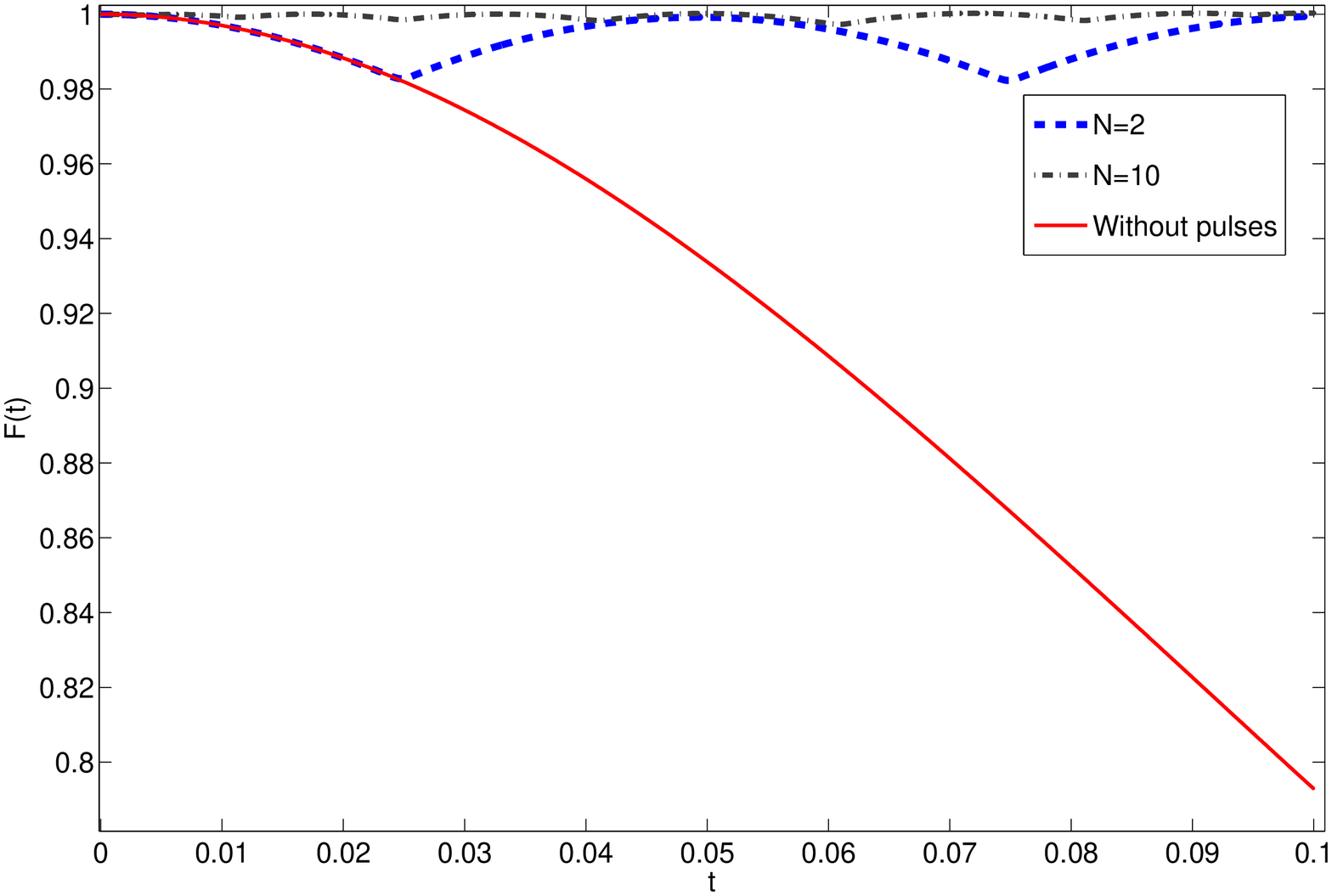}}
\end{center}
\par
 \caption{(color online) Expectation value of the coherence measure ${\cal P}_{|\Psi\rangle}$,
 denoted $F(t)$, as a function of time in dimensionless units, with $|\Psi\rangle$
 being one basis state of a three-level system. The central system
 is coupled with a bath modeled by other four three-level subsystems.
  The bottom curve represents significant decoherence without decoherence control.
  The top two curves represent decoherence suppression based on the control
  operator constructed in Eq. (\ref{control3}), for $N=2$ and $N=10$.}
\end{figure}

Finally, we show in Fig. 4 the decoherence suppression of a three-level quantum system,
with the control operator
given by Eq. (\ref{control3}).  Here the bath is modeled by other four three-level subsystems, and
the total Hamiltonian is chosen as
\begin{eqnarray}
 H & = & \sum_{m=2}^{5}\sum_{\alpha=\{x,y,z\}}b_{j,m}j_{\alpha, m}\nonumber \\
 &  & + \sum_{n=1}^{5}\sum_{\alpha=\{x,y,z\}}\sum_{m>n}^{5}\sum_{\beta=\{x,y,z\}}c_{\alpha\beta}j_{\alpha,m}j_{\beta,n}\nonumber \\
 &  & +\ H_{c},\end{eqnarray}
where $j_{\alpha,m}$ represents the $j_{x}$, $j_y$, or $j_z$
operator associated with the $m$th three-level subsystem, with the
first being the central system and the other four being the bath.
The coupling coefficients are again randomly chosen from $[0,1]$
with dimensionless units. The results are analogous to those seen in
Fig. 1 and Fig. 2, confirming the general applicability of our UDD
control sequence in multi-level quantum systems. Note also that even
for the $N=2$ case (middle curve in Fig. 4), decoherence suppression
already shows up clearly. The results here may motivate experimental
UDD studies using systems analogous to the kicked-top system
realized in Ref. \cite{Jessen}.

\section{Discussion and Conclusion}

So far we have assumed that the system-bath coupling, the bath self-Hamiltonian, and the system
Hamiltonian in the absence of the control sequence are all
time-independent. This assumption can be easily lifted.  Indeed, as
shown in a recent study by Pasini and Uhrig for single-qubit systems
\cite{Pasini}, the UDD result holds even after introducing a smooth time
dependence to these terms. The proof in Ref. \cite{Pasini} is also based on Yang
and Liu's work \cite{univUDD}.  A similar proof can be done for our
extension here.  Take the two-qubit case with the control operator
$Y_1$ as an example. If $H_0$ and $H'$ are time-dependent, then the
unitary evolution operator in Eq. (\ref{Ut}) is changed to
\begin{eqnarray}
U(T) & = & (-iY_1)^{N}\mathfrak{J}\left[ e^{-i\int _{T_N}^{T}[H_{0}+(-1)^{N} H'] dt}\right]  \nonumber \\
& &\ \times\ \mathfrak{J}\left[e^{-i\int_{T_{N-1}}^{T_N}[H_{0}+(-1)^{N-1}H']dt} \right] \nonumber \\
& &\ \cdots \nonumber \\
 & &\ \times\ \mathfrak{J}\left[ e^{-i\int_{T_2}^{T_3}[H_{0}+H']dt} \right]\nonumber \\
& &\ \times\ \mathfrak{J}\left[ e^{-i\int_{T1}^{T2} [H_{0}-H']dt} \right]\nonumber \\
& &\ \times\  \mathfrak{J}\left[e^{-i\int_{0}^{T_1}[H_{0}+H']dt}\right] \nonumber \\
&=& (-iY_1)^{N}\mathfrak{J} \left[e^{-i\int_{0}^{T}H_0 dt}\right] \mathfrak{J}\left[ e^{-i\int_{0}^{T}
F_{N}(t)H_{I}'(t)\mathrm{d}t}\right], \nonumber \\
\label{uorder}
\end{eqnarray}
with
\begin{eqnarray}
H_I'(t) = \mathfrak{J} \left[e^{i\int_{0}^{T}H_0 dt}\right] H' \mathfrak{J} \left[e^{-i\int_{0}^{T}H_0 dt}\right].
\end{eqnarray}
Because the term $\mathfrak{J} \left[e^{-i\int_{0}^{T}H_0 dt}\right]$ in Eq. (\ref{uorder}) does not affect
the expectation value of our coherence measure,  the final expression for the coherence measure
 is essentially the same as before and
is hence again given by its initial value multiplied by $1+O(T^{N+1})$.

%MM
Our construction of the UDD control sequence is based on a pre-determined coherence measure ${\cal P}_{|\Psi\rangle}$ that characterizes a certain type of quantum coherence.  This implies that our two-qubit UDD relies on which type of decoherence we wish to suppress.  Indeed,
this is a feature shared by Uhrig's work \cite{UDDprl} and the Yang-Liu universality proof \cite{univUDD} for single-qubit systems (i.e., suppressing either transverse decoherence or longitudinal population relaxation).
Can we also efficiently suppress decoherence of different types at the same time, or can we simultaneously preserve the quantum coherence associated with entangled states as well as non-entangled states?  This is a significant issue because the ultimate goal of decoherence suppression
is to suppress the decoherence of a completely unknown state and hence to preserve the quantum coherence of any type at the same time.
Fortunately, for single-qubit cases: (i) there are already good insights into the difference between decoherence suppression for a known state and decoherence suppression for an unknown state \cite{lorenza3,lorenza4} (with un-optimized DD schemes); and (ii)
a very recent study \cite{QDD} showed that suppressing the longitudinal decoherence
and the transverse decoherence of a single qubit at the same time in a ``near-optimal" fashion is possible, by arranging different control Hamiltonians in a nested loop structure. Inspired by these studies, we are now working on an extended scheme to achieve efficient decoherence suppression in two-qubit systems, such that two or even more types of coherence properties can be preserved.
Thanks to our explicit construction of the UDD control sequence for non-entangled and entangled states, some interesting progress towards this more ambitious goal is being made.  For example, we anticipate that it is possible to preserve two types of quantum coherence of a two-qubit state at the same time, if we have some partial knowledge of the initial state.

%MM
It is well known that decoherence effects on two-qubit entanglement
can be much different from that on single-qubit states.
 One current important topic is  the
 so-called ``entanglement sudden death" \cite{science}, i.e., how two-qubit entanglement can completely disappear
 within a finite duration.  Since the efficient preservation of two-qubit entangled states by UDD is
 already demonstrated here,
 it becomes certain that the dynamics of entanglement death can be strongly affected by applying just very few control pulses.
In this sense,
our results on two-qubit systems are not only of great experimental interest to quantum entanglement storage, but also of fundamental interest to understanding some aspects of entanglement dynamics in an environment.

To conclude, based on a generalized polarization operator as a
coherence measure, we have shown that UDD also applies to two-qubit
systems and even to arbitrary multi-level quantum systems. The
associated control fidelity is still given by $1+O(T^{N+1})$ if $N$
instantaneous control pulses are applied.
 This extension is completely general because no assumption on
the environment is made.  We have also explicitly constructed the
control Hamiltonian for a few examples, including a two-qubit system and
a three-level system. Our results are expected to advance both theoretical and
experimental studies of decoherence control.

\section{Acknowledgments}
This work was initiated by an ``SPS" project in the Faculty of
Science, National University of Singapore.  We thank  Chee Kong Lee
and Tzyh Haur Yang for discussions.
 J.G. is supported by the NUS start-up fund (Grant
No. R-144-050-193-101/133) and the NUS ``YIA" (Grant No.
R-144-000-195-101), both from the National University of Singapore.


\begin{thebibliography}{20}
\bibitem{BangBang1}L. Viola and S. Lloyd, Phys. Rev. A \textbf{58},
2733 (1998). \label{the: Bang-Bang DD}
\bibitem{BangBang2}L. Viola, E. Knill, and  S. Lloyd, \prl{\bf 82}, 2417 (1999).
\bibitem{exp2}J. J. L. Morton {\it et al}., Nature Physics {\bf 2}, 40 (2006);
J. J. L. Morton {\it et al}., Nature {\bf 455}, 1085 (2008).
\bibitem{UDDexact}G. S. Uhrig, New Journal of Physics \textbf{10}, 083024 (2008).
\bibitem{CDD1}K. Khodjasteh and D.A. Lidar, Phys. Rev. Lett. \textbf{95},
180501 (2005).
\bibitem{CDD2}K. Khodjasteh and D. A. Lidar, Phys. Rev. A \textbf{75},
062310 (2007). \label{the: CDD}
\bibitem{UDDprl}G. S. Uhrig, Phys. Rev. Lett. \textbf{98}, 100504
(2007).\label{the:UDD prl}
\bibitem{key-12}B. Lee, W. M. Witzel, S. Das Sarma, Phys. Rev. Lett.
\textbf{100}, 160505 (2008).\label{the: UDD-universal the idea}
\bibitem{UDDvsCPMG}M. J. Biercuk, H. Uys, A. P. VanDevender, N. Shiga,
W. M. Itano, and J. J. Bolinger, Nature \textbf{458}, 996 (2009).
\bibitem{detailpra}M. J. Biercuk, H. Uys, A. P. VanDevender, N. Shiga,
W. M. Itano, and J. J. Bolinger, \pra{\bf 79}, 062324 (2009).
\bibitem{Du-Liu} J. F. Du, X. Rong, N. Zhao, Y. Wang, J. H. Yang, and R. B. Liu, Nature {\bf 461}, 1265 (2009).
\bibitem{univUDD}W. Yang and R. B. Liu, Phys. Rev. Lett. \textbf{101},
180403 (2008).
\bibitem{Jessen}S. Chaudhury, A. Smith, B. E. Andersdon, S. Ghose, and P. S. Jessen,
Nature {\bf 461}, 768 (2009).
\bibitem{haake} F. Haake, {\it Quantum Signatures of Chaos} 2nd Ed. (Springer-Verlag,
Berlin, 1999).
\bibitem{Gongprl} See, for example,  J. Wang and J. B. Gong, Phys. Rev. Lett. {\bf
102}, 244102 (2009) for an extensive list of the kicked-top model
literature and its relevance to several areas.
\bibitem{Pasini}S. Pasini and G. S. Uhrig, arXiv:0910.0417 (2009).
\label{the: UDD-universality time dependent}
\bibitem{lorenza3}W. X. Zhang, N. P. Konstantinidis, V. V. Dobrovitski, B. N. Harmon, L. F. Santos, and L. Viola,
\prb{\bf 77}, 125336 (2008).
\bibitem{lorenza4}W. X. Zhang, V. V. Dobrovitski, L. F. Santos, L. Viola, and B.N. Harmon, \prb{\bf 75}, 201302 (R) (2007).

\bibitem{QDD}J. R. West, B. H. Fong, and D. A. Lidar, arXiv:0908.4490v2
(2009).

\bibitem{science}T. Yu and J. H. Eberly, Science {\bf 323}, 598 (2009).



\end{thebibliography}
\end{document}